\documentclass[aps,prl,twocolumn,showpacs]{revtex4}

\usepackage{amsmath}
\usepackage{graphicx}

\allowdisplaybreaks

\def\be{\begin{equation}}
\def\ee{\end{equation}}
\def\bea{\begin{eqnarray}}
\def\eea{\end{eqnarray}}

\begin{document}

\title{Creation of a molecular condensate by dynamically melting a Mott-insulator}

\author{D.~Jaksch,$^1$ V. Venturi,$^3$ J.I.~Cirac,$^2$ C.J.~Williams,$^3$ and P.~Zoller$^1$}

\affiliation{${}^1$ Institute for Theoretical Physics, University
of Innsbruck, A--6020 Innsbruck, Austria.}

\affiliation{${}^2$ Max-Planck Institut f\"ur Quantenoptik,
Hans-Kopfermann Str. 1, D-85748 Garching, Germany.}

\affiliation{${}^3$Atomic Physics Division, National Institute of Standards and Technology,
Gaithersburg, Maryland 20899-8423.}

\begin{abstract}
We propose creation of a molecular Bose-Einstein condensate (BEC)
by loading an atomic BEC into an optical lattice and driving it
into a Mott insulator (MI) with exactly two atoms per site.
Molecules in a MI state are then created under well defined
conditions by photoassociation with essentially unit efficiency.
Finally, the MI is melted and a superfluid state of the molecules
is created. We study the dynamics of this process and
photoassociation of tightly trapped atoms.
\end{abstract}


\maketitle

The generation of Bose Einstein condensates (BEC) of dilute atomic
gases has resulted in a remarkable series of experiments
demonstrating various properties of quantum degenerate gases
\cite{BEC}. One of the next major goals in this effort is the
realization of a molecular BEC. A promising route towards a
molecular condensate is the conversion of an atomic BEC to
molecules via photoassociation, a process discussed so far for
conditions of quasihomogeneous trapping of atomic gases
\cite{Heinzen,Javanainen,Holland,TwoColor}.  In this Letter we
describe a novel path to create condensates of composite atomic
objects, in particular a molecular BEC, based on photoassociation
via a {\em Mott insulator state} of bosonic atoms trapped in an
optical lattice \cite{Mott,MottExp}. This provides an efficient
way of generating a molecular BEC, avoiding some of the problems
encountered in the quasihomogeneous case \cite{Heinzen}. It also
touches upon fundamental questions related to the formation of a
BEC by ``melting'' of a Mott-insulator (MI) state in a quantum
phase transition, as opposed to the familiar growth from a thermal
cloud of atoms \cite{growth}.

Experimental advances in manipulating BECs \cite{BEC}, and in
particular the loading of a BEC into an optical lattice generated
by interfering laser beams have recently led to a seminal
experiment by  I.~Bloch and collaborators \cite{MottExp}. They
demonstrated a quantum phase transition from a BEC or superfluid
(SF) state into a MI by varying the lattice laser intensity, as
proposed theoretically in \cite{Mott}. While a SF phase has long
range order, the MI phase corresponds to the loading of a precise
number of atoms into each lattice site, i.e. Fock state occupation
of lattice sites. Among the proposed applications of this new
atomic quantum phase are the study of ultracold controlled
collisions and quantum computing with neutral atoms \cite{CCC}. In
the present context, the MI phase opens the possibility to
efficiently create a molecular BEC in the following four steps:
(i) an atomic BEC is loaded into an optical lattice, (ii) the
depth $V_0$ of the optical lattice is increased to create a MI
with {\em exactly two particles per lattice site}, (iii) a
molecular MI state is produced by two-color photoassociation of
the atoms under tight trapping conditions, and (iv) by decreasing
the depth of the optical lattice the MI state is ``melted'', and
thus a molecular BEC is created in a quantum phase transition.

At the end of step (i) above, we have an ensemble of bosonic atoms
illuminated by orthogonal, standing wave laser fields tuned far
from atomic resonance. These laser fields generate a potential for
atomic motion of the form $V(\vec x) = \sum_{i=1}^{3} V_{0i} {\rm
sin}^2(k x_i)$ with $k=2\pi / \lambda$ the wave-vector of the
light and lattice period $a=\lambda/2$.  The dynamics of bosonic
atoms occupying the lowest Bloch band of an optical lattice is
well described by the Bose-Hubbard model (BHM) \cite{Mott} which
includes the interaction $U_a$ between particles occupying the
same lattice site and the tunneling $J_a$ of particles from one
site to the next. The BHM Hamiltonian is given by
\begin{equation}
H_a=-J_a\sum_{\langle i,j \rangle}a_{i}^{\dagger
}a_{j}+\sum_{i}\epsilon_{i}\hat{n}_{i} \, + \frac{1}{2}U_a
\sum_{i}\hat{n}_{i}(\hat{n}_{i}-1)  \label{BH}
\end{equation}
where $a_i$ is a bosonic destruction operator of a particle at
site $i$. The number of particles in site $i$ is given by the
operator $\hat{n}_i=a_i^\dagger a_i$, and $\epsilon_i$ is an
energy offset due to an external trapping potential.  The first
sum in Eq.~(\ref{BH}) runs over all nearest neighbors denoted by
$\langle i,j \rangle$. Increasing the laser intensity of the
trapping laser tends to compress atoms near the nodes of the
lattice field, and thus leads to an increased on-site interaction
$U_a$, while the atomic tunneling rate $J_a$ decreases
\cite{Mott}. The BHM predicts a quantum phase transition from the
SF phase to the MI state: according to mean field theory this
occurs at the critical value $U_a^{(c)} \approx 5.8 z J_a$
\cite{MottOld} with $z$ the number of nearest neighbors of each
site. This corresponds to a (moderate) potential depth of $V_0=10
E_R$ where $E_R = \hbar^2 k^2/2m$ is the recoil energy for atoms
with mass $m$.

\begin{figure}[tbp]
\begin{center}
\includegraphics[width=0.9\linewidth]{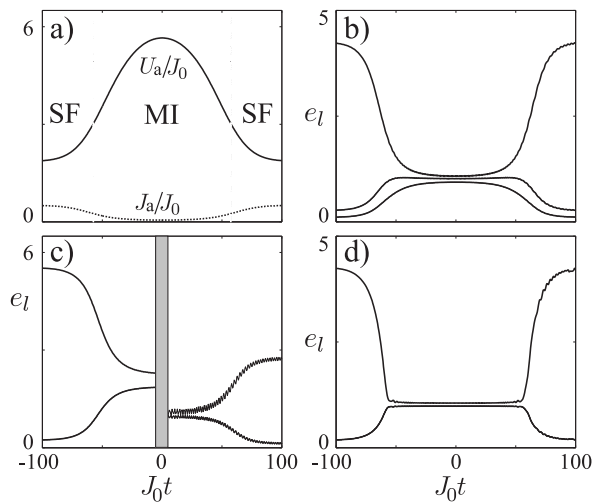}
\end{center}
\caption{a) Time dependence of $U_a$ (solid curve) and $J_a$
(dotted curve). The vertical dashed
lines separate the SF and MI regions expected for adiabatic time
evolution of the system in 1D. Parameters: we assume
${}^{87}$Rb and $\lambda = 390$nm with $
V_0(t)=V_{\rm SF} + (V_{\rm MI}-V_{\rm SF})/ (1+\exp((t^2-t_w^2)/
t_s^2))$, $t_w=30/J_0$, $t_s=40/J_0$, $V_{\rm MI}=20 E_R$,
$V_{\rm SF}= 5 E_R$ giving $J_0=0.13 E_R$. $J_0/z$ is the hopping
matrix element for $t \rightarrow - \infty$
 b) Transition from the SF to the MI back
to the SF phase for atoms. We plot $e_l$ (lower two curves are
doubly degenerate) against $t$ for $N=M=5$ in a 1D lattice with
periodic boundary conditions. c) Atoms in the SF phase are driven
to a MI phase (time interval $t<0$), converted to a molecular MI
phase by a Raman pulse (shaded region around $t=0$), and melted to
obtain a molecular BEC ($t>0$). We plot $e_l$ (lower curve doubly
degenerate) for the atoms (molecules) before (after) the
conversion for $N/2=M=3$ in 1D, and parameters $J_b=J_a/2$,
$U_a=U_b=U_{ab}$ with time dependence given in Fig.~\ref{fig1}a).
d) Same as b) but using the Gutzwiller ansatz (the lower curve is
four fold degenerate). \label{fig1}}
\end{figure}

We will first illustrate the dynamics of the BHM Hamiltonian with
a time dependent depth $V_0(t)$ of the optical lattice controlled
by the laser intensity, leading to a variation $U_a(t)$ and
$J_a(t)$ in Eq.~\ref{BH}. We assume the system initially to be in
the SF ground state and calculate its time evolution for the time
dependence shown in Fig.~\ref{fig1}a \cite{Mott}. For a model
problem of $N$ particles, where $N$ is small ($\approx 10$), in a
few lattice sites, the time dependent Schr{\"o}dinger equation for
the wave function $\Psi(t)$ can be solved exactly.
Fig.~\ref{fig1}b plots the eigenvalues $e_l$ of the one particle
density matrix $\rho_{i,j}= \langle a_i^\dagger a_j \rangle$
($\equiv \langle \Psi(t)|a_i^\dagger a_j |\Psi(t) \rangle$). As
expected, there is one large eigenvalue for $U_a < U_a^{(c)}$ of
the order of the number of particles. The corresponding wave
function is approximately given by $|\psi_{\rm SF}\rangle \propto
(\sum_i a_i^\dagger)^N |{\rm vac} \rangle$ for $N$ particles in
$M$ sites with $|{\rm vac}\rangle$ the vacuum state. All the other
eigenvalues are small and are associated with the quantum
depletion of the SF state. As $U_a$ increases and crosses the
critical point $U_a^{(c)}$ all eigenvalues tend towards one,
corresponding to a diagonal single particle density operator (MI).
Upon ramping $U_a$ down again the SF is restored.

\begin{figure}[tbp]
\begin{center}
\includegraphics[width=0.8\linewidth]{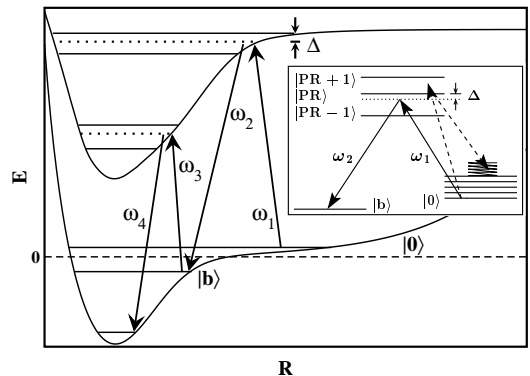}
\end{center}
\caption{Production of ground state molecules using two Raman
transitions. The inset shows the first Raman transition (solid
lines) and the process where $\omega_2$ is absorbed before
$\omega_1$ (dashed lines). For $^{87}$Rb and for a trap frequency
of $\nu=1$MHz, $E_{0}\approx2$ MHz and outer turning point
$R_{TP}\approx500 \, a_0$ ($a_0$=0.0529nm). For the singlet
pathway, $E_{b}\approx -32$ GHz and $R_{TP}\approx 32 \, a_0$.
\label{fig3}}
\end{figure}

An extension of the BHM (\ref{BH}) in an optical lattice describes
the situation with atoms and molecules present.  Denoting the
annihilation operator of a molecule by $b_i$, we add to the
Hamiltonian (\ref{BH}) a tunneling $J_b$ and on-site interaction
term $U_b$ for molecules, and an atom-molecule interaction term of
the form $H_{a-b}= U_{ab} \sum_i b_i^\dagger b_i a_i^\dagger a_i$.
The underlying assumption is that the laser beams
generate an optical lattice for atoms and molecules with the
same structure of nodes and antinodes, although both lattices can
have different depth. The process of atom-molecule conversion by
photoassociation is described by the Hamiltonian $H_{\rm conv} =
\Omega(t)  \sum_i \left( b_i^\dagger a_i a_i + {\rm h.c.}\right) /
\sqrt{2} $, where $\Omega(t)$ is an effective Rabi frequency which
is turned on at time $t=0$ for a short time interval to convert
two atoms at one lattice site into one molecule. In practice, this
process will consist of several Raman steps to go to the molecular
ground state (see Fig.~(\ref{fig3})).

We emphasize several distinguishing key features of the
atom-molecule conversion process in an optical lattice.  First
conversion is most efficient under tight trapping conditions (in a
regime where tunneling between lattice sites is negligible). The
high atomic densities associated with the strong compression of
the atoms opens inelastic collision channels that typically quench
all lattice sites with three or more atoms. Thus we assume that
only lattice sites with occupation of two (or one) atom survive
the lattice compression. These loss processes are added to our
Hubbard dynamics by writing down a master equation for a density
matrix that includes these decay channels. However, since in the
MI phase atomic number fluctuations are small, there are only a
very few lattice sites that are actually depleted by this loss
process. Second, for (exactly) two atoms trapped at one lattice
site we have a complete microscopic understanding of the two-atom
dynamics, and the conversion to molecules by photoassociation
\cite{Mies} beyond the effective description contained in the
Hubbard model.

The two-atom Schr{\"o}dinger equation \cite{Mies} at a given
lattice site separates for harmonic confinement into
center-of-mass and relative coordinates. The potential curves for
the relative motion of the two atoms in the trap are schematically
shown in Fig.~\ref{fig3}: for small distances $R$ we have the
familiar molecular Born-Oppenheimer potentials while the large
$R$-behavior is dominated by the trap confinement. The trapping
potential discretizes the molecular continuum (scattering) states
to form a series of harmonic oscillator trap states with frequency
$\nu = (4 V_0 E_R)^{1/2}/\hbar$ with $V_0$ the potential depth.
The goal is to perform a Rabi flop $\Omega T=\pi$ from the lowest
trap state of two atoms (i.e. the state associated with the lowest
atomic Bloch band) to a bound molecular state, with $\Omega$ the
two-photon Rabi frequency.  The discreteness of the trap states
makes this a bound-bound Raman transition. The condition for not
exciting any other trap states, i.e. to avoid heating, is $\Omega
\ll \nu$. On the other hand, for incoherent processes (as
spontaneous decay from the intermediate state) with effective
decay rate $\gamma$ to be small, we must have $\gamma T \ll 1$.
Thus we require a large two-photon Rabi frequency and tight
trapping, $\gamma \ll \Omega \ll \nu$. Note that $\Omega$ involves
a matrix element from the bound trap state to a molecular state,
which results in a scaling $\Omega \propto \nu^{3/4}$ \cite{Mies},
i.e.~under conditions of tight trapping the two-photon Rabi
frequency will be significantly enhanced. We will give specific
numbers for these parameters for the case of ${}^{87}$Rb below.
Thus it is the preparation of the two-atom MI phase together with
strong confinement which guarantees the coherent conversion of
atoms to molecules with essentially unit efficiency.

Fig.~\ref{fig1}c shows results from the exact integration of the
Schr{\"o}dinger equation for three sites with six atoms. Starting
from the SF phase, the atoms are driven to the MI phase and
converted into molecules at $t=0$. Melting of the molecular MI
phase then produces a molecular SF.

To describe the dynamics with a large number of particles in 2D
and 3D we employ a time dependent mean field approximation based
on a Gutzwiller ansatz \cite{MottOld}. For simplicity of writing
we consider for the moment the case of atoms alone, where we write
the wave function as the product of superposition states at the
various lattice sites, $ |G(t)\rangle =\prod_{i=1}^{M}\left(
\sum_{n=0}^ {\infty} f_{n}^{(i)}(t)|n\rangle_{i}\right) $. This
ansatz is motivated by the success and simplicity of
time-independent Gutzwiller mean field theory to model the ground
state and phase diagram of the BHM \cite{MottOld}. The ground
state is obtained from the variational principle $\langle
G|H|G\rangle -\mu \langle G|\hat{N} |G\rangle \rightarrow \min $,
where $\mu $ is a chemical potential introduced to enforce a given
mean particle number \cite{Mott}. From the time-dependent
variational principle, $\langle G(t)|i\hbar \frac{\partial
}{\partial t}-H(t)|G(t)\rangle \rightarrow \min$, the following
time dependent equation is readily derived:
\begin{eqnarray}
i\dot{f}_{n}^{(i)}&=&\frac{U_{a}}{2}n(n-1)f_{n}^{(i)}  \label{G} \\
&&-J_{a}\sum_{\langle
i,j\rangle }\left( \Phi _{j}^{\ast }f_{n+1}^{(i)}\sqrt{n+1}+\Phi
_{j}f_{n-1}^{(i)}\sqrt{n}\right), \nonumber
\end{eqnarray}
where $\Phi_{i} \equiv \langle G|a_{i}|G\rangle =\sum_{n}
f_{n-1}^{(i)*}\sqrt{n}f_{n}^{(i)}$ is the atomic SF density.
Eq.~(\ref{G}) is a nonlinear equation for the amplitudes
$f_{l}^{(i)}$, which preserves both normalization of the wave
function and the mean particle number. In the SF limit and for a
coherent state distribution of
$f_{n}^{(i)}=\psi_{i}^{n}\exp(-\left| \psi_{i} \right|^{2}/2) /
\sqrt{n!}$, Eq.~(\ref{G}) reduces to a time-dependent
Gross-Pitaevskii equation for $\psi _{i}$ on a lattice.

By projection to a state with definite particle number $N$,
$|G_{N}\rangle = \mathcal{P}_{N}|G\rangle /\left\Vert
\mathcal{P}_{N}|G\rangle \right\Vert \sim \int_{0}^{2\pi} d\varphi
\exp(iN\varphi) \prod_{i=1}^{M} (\sum_{n=0}^\infty \exp
(-in\varphi) f_{n}^{(i)}|n\rangle_{i})$ a more consistent
description is obtained. The resulting time dependent
Schr\"{o}dinger equation for the amplitudes $f_{n}^{(i)}$ is
significantly more complex. It can be shown, however, that
$f_{n}^{(i)}$ of the number projected Gutzwiller wave function
again obey Eq.~(\ref{G}), provided the variance of the particle
numbers at each lattice site satisfies $\Delta n_{i}\gg
1/\sqrt{N}$ (where $\Delta n_{i}^{2}\equiv \langle
n_{i}^{2}\rangle -\langle n_{i}\rangle ^{2}\ $ with $\langle
n_{i}^{2}\rangle =\sum_{n}n^{2}|f_{n}^{(i)}|^{2}$ and $\langle
n_{i}\rangle =\sum_{n}n|f_{n}^{(i)}|^{2}$). Note that this
excludes the regime where we have a \emph{precise} locking of the
particle number, i.e., $f_{n}^{(i)}=\delta _{n,n_{0}}$, as in the
MI obtained from the non-number conserving Gutzwiller for a
homogeneous situation, when the number of particles $N$ is
commensurate with the lattice sites $M$ and $n_0=N/M$. However, a
precise Fock state is never realized in the time evolution we
consider since the initial superfluid density is not completely
destroyed while ramping the optical lattice up. Also, an
additional trap potential confining the system to a certain region
in space ensures the existence of a remnant superfluid component.
Below we model the evolution of an initial SF to an (approximate)
MI while changing $U_a$ and $J_a$ by integrating mean field
equations of the type (\ref{G}). Fig.~\ref{fig1}d gives the
results obtained from Gutzwiller theory with initial state given
by the time-independent Gutzwiller wave function, to be compared
with the exact integration of the time dependent Schr\"{o}dinger
equation in $1$D for a few particles in Fig.~\ref{fig1}b. As
expected, mean field theory shows a more pronounced phase
transition than the few atom $1$D calculation.

\begin{figure}[tbp]
\begin{center}
\includegraphics{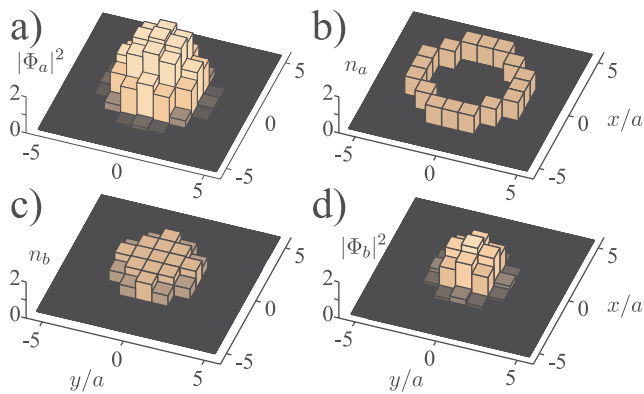}
\end{center}
\caption{2D lattice. a) Initial atomic SF density $|\Phi_a|^2$. b)
Number of atoms $n_a$  and  c) number of molecules $n_b$
immediately after Raman conversion. d) Final SF molecular density
$|\Phi_b|^2$. The chemical potential is $\mu=2.5 J_0$, the
additional trap potential is given by $\epsilon_i= 2 J_0
(x_i^2+y_i^2)/5 a^2$, where $x_i,y_i$ denote the coordinates of
well $i$.  The depth of the optical lattice is changed $10$ times
slower than in Fig.~\ref{fig1}. \label{fig2}}
\end{figure}

Using a generalization of the Gutzwiller ansatz to include
superposition states of atoms and molecules, $|G(t)\rangle
=\prod_{i=1}^{M}\left( \sum_{n_a,n_b=0}^{\infty} f_{n_a,
n_b}^{(i)}(t) |n_a, n_b\rangle_{i} \right)$, where $n_a$ and $n_b$
refer to the atomic and molecular occupation, respectively, we
have numerically investigated the creation of a molecular BEC in a
2D lattice with a superimposed harmonic trapping potential. The
results are shown in Fig.~\ref{fig2}. As expected we find that
molecules are only created in sites with an atomic occupation of
two before the Raman process (cf.~Fig.~\ref{fig2}b). They are
surrounded by a ring of atoms which originates from those site
with an atomic occupation of one before the Raman transition (see
Fig.~\ref{fig2}c). Finally, Fig.~\ref{fig2}d shows the superfluid
molecular density after ramping the optical lattice down.

We now turn to the description of the coherent Raman transitions
involved in creating molecules in step (iii) of our scheme. There
are several constraints on the choice of detunings and laser
intensities (see Fig.~\ref{fig3}) to maximize $\Omega$. First, the
detuning $\Delta$ from an allowed excited photoassociation
resonance $|\Psi_{\rm PR} \rangle$ (cf.~Fig.~\ref{fig3}) must be
large compared to the natural linewidth $\gamma_{PR}$ of the
intermediate state to suppress spontaneous Raman scattering. Also,
one should not detune half-way between two resonances since
interference between these two intermediate states will lead to a
minimum in the effective two-photon Rabi frequency $\Omega$. For
appropriate choices of $\Delta$ and the intermediate state
$|\Psi_{\rm PR} \rangle$, the coupling to all intermediate states
other than $|\Psi_{\rm PR} \rangle$ can be neglected. Second, as
outlined above, we have to ensure that the process where a
$\omega_2$ photon is absorbed before a $\omega_1$ photon
(schematically shown in the inset of Fig.~\ref{fig3}) has
negligible probability. This process causes trap excitations of
single atoms and thus leads to heating. If both of these
conditions are fulfilled the effective Rabi frequency $\Omega$ on
Raman resonance for the first Raman step is given by $\Omega =
\Omega_1 \Omega_2 / 2 \Delta$, where $\Omega_{1,2}$ are the Rabi
frequencies for the first (second) step, and the effective
spontaneous emission rate is $\gamma = \gamma_{\rm PR}
\Omega_1^2/4\Delta^2$.

For the case of $^{87}$Rb, the following Raman pathway is viable
for producing $X^1 \Sigma_g^+ (v=0,J=0)$ molecules:
\begin{eqnarray}
&& X^1 \Sigma_g^+ (v_{\mathrm{trap}}=0)\rightarrow
A^1 \Sigma_u (v=213) \rightarrow
X^1 \Sigma_g^+ (v=120)  \nonumber \\
&& X^1 \Sigma_g^+ (v=120)  \rightarrow
A^1 \Sigma_u (v=185) \rightarrow
X^1 \Sigma_g^+ (v=52) \nonumber  \\
&& X^1 \Sigma_g^1 (v=52) \rightarrow
A^1 \Sigma_u (v=24) \rightarrow
X^1 \Sigma_g^+ (v=0)  .
\label{path}
\end{eqnarray}
A two-step Raman pathway also exists for producing $a^3 \Sigma_u^+
(v=0,J=0)$ molecules. For the first step of the singlet pathway,
the vibrational spacing near the $A^1 \Sigma_u (v=213)$ level is
110 GHz, $\gamma_{PR} \approx 12$ MHz, and the binding energy of
the $X^1 \Sigma_g^+ (v=120)$ is 31.9 GHz. Given a trap frequency
$\nu$ of 1 MHz and intensities $I_1=1 \mbox{W}/\mbox{cm}^2$ and
$I_2=10^{-3} \mbox{W}/\mbox{cm}^2$ for the first Raman step of
Eq.~(\ref{path}), one obtains $\Omega_1=0.71$ MHz and
$\Omega_2=3.7$ MHz. For a red detuning of 200 linewidths we get
$\Omega=1.1$ kHz and and an effective spontaneous Raman scattering
rate of $\gamma=5.5$ Hz. For the process described in the
preceding paragraph one finds that the undesirable
$\omega_2-\omega_1$ process has a blue detuning of $29.5$ GHz off
the same intermediate level and a Rabi frequency that is more than
a factor of 100 below $\Omega$. Although the pathway for producing
singlet molecules is partially optimized, no attempt to optimize
the intensities and detunings has been made. However we have
chosen values to show that the undesirable $\omega_2-\omega_1$
process can be sufficiently suppressed. The rate limiting step in
the overall pathway is determined by the matrix element of the
first Raman step. The matrix elements for all subsequent steps are
at least 3-orders of magnitude larger than this one.

For experimental purposes, molecules in higher lying vibrational
levels of the $X^1 \Sigma_g^+$ or $a^3\Sigma_u^+$ state should
also allow the formation and detection of a molecular BEC.
Vibrational relaxation caused by inelastic molecule-molecule
collisions, that ultimately limits the lifetime of the resulting
molecular condensate, will in general be strongly suppressed as
the kinetic energy in the exit channels increases. This occurs as
the vibrational spacing increases and should allow for
observations of molecular BEC for moderately bound vibrational
levels.

In conclusion, generation of a MI phase of atoms allows efficient
conversion of atoms to molecules, and to obtain a molecular
condensate via ``melting'' in a quantum phase transition. This
idea can be immediately generalized to  e.g. heteronuclear
molecules, or one could use laser chemistry to build more complex
composite objects (trimers etc.) and, possibly, corresponding
condensates by quantum melting. The key to designing these
processes is the fact that the MI phase provides us with a given
small number of particles per lattice site (reaction partners)
whose few body dynamics can be understood in all detail and
controlled via laser interactions.

Discussions with E.~Tiesinga, P.S.~Julienne and R.~Grimm are
gratefully acknowledged. Work supported in part by the Austrian
Science Foundation, EU Networks, and by the U.S. Office of Naval
Research.

\end{document}